\documentstyle[epsfig,12pt]{article}
\begin{document}

\begin{center}
{\Large Non-Linear Dynamics in Patients with Stable Angina Pectoris
\footnote{Presented at Computers in Cardiology, September 23-26, 2001 -
Rotterdam, The Netherlands}} \\
\vspace*{0.3cm}
{\large G. Krsta\v ci\' c$^{a}$, M. Martinis$^{b}$, E. Vargovi\' c$^{c}$, 
A. Kne\v zevi\' c$^{b}$, A. Krsta\v ci\' c$^{d}$} \\
\vspace{0.3cm}
Institute for Cardiovascular Disease and Rehabilitation$^{a}$, "Rudjer
Bo\v skovi\' c" Institute$^{b}$,
CDV info$^{c}$, University Clinic "Vuk Vrhovac"$^{d}$,
 Zagreb, Croatia
\end{center}

\begin{abstract}

Dynamic analysis techniques may quantify abnormalities in heart rate
behaviour based on non-linear and fractal analysis.The applicability of
these new methods for detecting complex heart rate dynamics in coronary
heart disease is not well established.
To investigate the clinical and prognostic significance of fractal
dimension and detrended fluctuation analysis, two different groups were
studied, group of patients with stable angina pectoris without previous
myocardial infarction, and age-matched healthy controls.
The fractal dimension of the RR series was determined using the rescaled
range (R/S) analysis technique. To quantify fractal
longe-range-correlation properties of heart rate variability, the
detrended fluctuation analysis (DFA) technique was used. The heart rate
variability was characterized by a scaling exponent $\alpha$, separately for
short-term ($<$ 11 beats) and for long-term ($>$ 11 beats) time scales.
The results of data sets show the existence of crossover phenomena between
short-time scales. In patients with stable angina pectoris the short-term 
fractal scaling exponent ($\alpha_{1}$) was significantly lower 
(0.95 $\pm$ 0.05 vs. 1.08 $\pm$ 0.06; p $<$ 0.05), 
while there were no differences in the long-term
fractal scaling exponent ($\alpha_{2}$)(1.37 $\pm$ 0.26 vs. 1.39 $\pm$ 0.04; 
p$<$0.05). The
patients with stable angina pectoris also had higher fractal dimension
than a healthy control group (p $<$ 0.05). The short-time scaling exponent
and fractal dimension are better than other heart rate variability
parametars in  differentiating patients with stable angina pectoris from
healthy subjects. Dynamic analysis may thus complement traditional
analysis in detecting altered heart rate behaviour.

\end{abstract}

\section{Introduction}

Heart rate variability (HRV) \cite{ref1,ref2} reflects the modulation of 
the cardiac function
by autonomic and other physiological systems, and its measurements from
ambulantory electrocardiography (ECG) recordings during an exercise ECG
test may be a useful method for both clinical and scientific purposes.

Traditional linear statistical measures provide limited information
about HRV, because non-linear mechanisms may also be involved in the
genesis of HR dynamics. A number of new methods have recently been
developed to quantify complex heart rate dynamics. They may uncover
abnormalities in the time series data, which are not apparent when using
conventional linear statistical methods.
This study tested the hypothesis that fractal measurements of HRV are
altered in patients with stable angina pectoris \cite{ref3,ref4}.

\section{Methods}

\subsection{Patients}

Twenty five consecutive patients with stable angina pectoris and
without previous myocardial infarction were included in the analysis
after the history of chest pain and non-invasive cardiovascular diagnostic
measurements (ECG at rest, echocardiography, 24 hours ECG,
vectorcardiography, exercise ECG test  and laboratory coronary risk
factors measurements) with ECG evidence of ischemic STsegment
depression ($>$ 0.1 mV) during an exercise test.
They were 57 $\pm$ 6 years old, 12 male. No cardiac medication was allowed
on the day of testing, and b - blocking therapy was withdrawn at least
7 days before and calcium antagonists at least 2 days before.
Patients with anginal chest pain, silent ischemic ST-segment depression
during the 24 hour ECG recording and diabetes mellitus were excluded.
The control group consisted of 20 randomly selected age matched (mean
age 58 $\pm$ 8 years), and sex matched (11 male) healthy subjects.
After a complete non-invasive eexamination and their medical
history all patients revealed no cardiovascular disease or use of medication. 
All controls had normal ECG at rest, echocardiographic data (M-mode, 2D
dimensional and Doppler echocardiography), normal arterial blood
pressure and fasting blood glucose. Subjects with evidence of ischemic
ST-segment depression during the exercise ECG test or the 24 hour ECG
recording were not included.
An exercise ECG on all subjects was obtained using a symptom or ECG
changes limited test, increasing the workload in a controlled manner. A
horizontal or downsloping ST-segment depression of $>$  0. mV occurring
0.08 seconds after the J point was considered to be of ischemic origin.

\subsection{Analysis of HRV}

Series of RR intervals were obtained from high resolution ECG (sampling
frequency 1000 Hz), and the recording time scale during the exercise ECG
test was approximately about 1500 beats. The ECG data were digitized by
the Vawebook 512 (Iotech. Cal. USA), and transferred to a computer for
analysis.
The RR interval series was passed through a filter that eliminated
noise, artifacts and premature beats. All RR interval series were first
edited automatically, after which careful manual editing was performed
by visual inspection of each RR interval. After this, all
questionable portions were excluded manually, and only segments with $>$
85 \% sinus beats were included in the final analysis.

The fractal dimension of the RR interval series was determined by the
"Rescaled range" (R/S) analysis \cite{ref5}: 
$R(n)/S(n) \sim n^{H}$, where
H is the Hurst's exponent, an important parameter used to characterize
the time series. 
$H \sim log (R/S)/log(n)$ where n is the length of the time box.

If the Hurst exponent is approximately about 0.5, it represents an
ordinary random walk or Brownian motion. In the cases where H is $<$ 0.5, it
means negative correlation between the increments (antipersistent time
series) and if it is $>$ 0.5, it means positive correlation between the
increments (persistent time series which are plentiful in nature).

The Hurst exponent is related to the fractal dimension (FD): H = E + 1 + –
FD, where E is the Euclidean.dimension ( E = 0 for point, 1 for line and
2 for surface). The relation between H and FD of the graph of a random
fractal is FD = 2H for one  dimensional signal. While H vary from 0 to 1, 
FD decreases from 2 to 1.

The Hursts exponent as well as  the fractal dimension were determined for
the whole time series during the exercise ECG, and also separately for
each program of exercise including half a minute baseline ECG before the
exercise and six minutes of relaxation after the exercise (Figs.1 and
2).

To quantify fractal long-range correlation properties of HRV, the
detrended fluctuation analysis (DFA) technique \cite{ref6}, 
which is a modified
root-mean-square (rms) analysis of a random walk, was used. The method
quantifies the presence or absence of fractal long-range correlation
properties. The rms fluctuation of integrated and detrended
time series is calculated by the formula
$$ F(n)= \sqrt{\frac{1}{N} \sum_{k=1}^{N}[y(k)-y_{n}(k)]^{2}} \, \, \, , $$
where \[ y(k)=\sum_{i=1}^{k} (R(i)-{\overline R}) \] and $y_n(k)$ is the
regression line through the data y(i) in the box of length n.
 This computation is repeated over all time scales (box sizes n)  in
order to characterize the relationship between F(n), the average
fluctuation, as a function of box size. Typically, F (n) will increase
with box size n. A linear relationship in a log--log plot indicates  the
presence of power law (fractal) scaling.
In this study, HRV was characterized by a scaling exponent $\alpha$, the slope
of the linear relationship between log F(n)  and log n, separately for
short-term ($<$ 11 beats, $\alpha_{1}$ ), and long-term ($>$ 11 beats,
$\alpha_{2}$)
fluctuations in the RR series data  (Fig.3).

\subsection{Statistical analysis}

Results are expressed as mean $\pm$ standard deviation (SD). The p value $<$
0.05 was considered significant.

\section{Results}

 The baseline clinical and heart rate variables of healthy controls and
patients with stable angina pectoris are listed in Table 1. There were
no differences observed in conventional statistical linear measures of
the HRV (average RR intervals and SDNN in the time domain and LF/HF ratio in
frequent domain). The results of exercise data sets show the existence of
crossover phenomena between short-time scales when using the DFA method.
A significant difference was found between patients with stable angina
pectoris and healthy controls in short-time scales (0.95 $\pm$ 0.05 vs. 1.08
$\pm$ 0.06) as can be seen in Figure 3. However there were no significant
differences in long-term series.  The fractal dimension was
significantly higher in patients with stable angina pectoris.

\begin{table}
\begin{tabular}{lll}
\hline
Clinical data (n=45): & Healthy Controls (n=20) & Patients with SAP (n= 25)\\
\hline
Age (yrs) &  58 $\pm$ 8 &  57 $\pm$ 6 \\
Men/women & 11/9  &  12/13 \\
ECG at rest (freq.) &  74 &  81 \\
VPCs/hour   &  3 $\pm$ 0.7  &    4 $\pm$ 2.7 \\
LV ejection fraction(\%) & 71 $\pm$ 6 & 63  $\pm$ 9 \\
E/A wave   &  1.4 $\pm$ 0.2  &  0.8 $\pm$ 0.2 \\
Exercise ECG data:   &  &  \\
Average RR interval (ms)  &  874 $\pm$ 108  &  856 $\pm$ 114 \\
SDNN (ms) &  149 $\pm$ 41 & 139 $\pm$ 40 \\
LF/HF ratio  &  3.2 $\pm$ 1.5 & 3.5 $\pm$ 1.7 \\
Hurst's exponent   &  0.81 $\pm$ 0.05 &  0.64 $\pm$ 0.07 * \\
Fractal dimension  &   1.19 $\pm$ 0.07 & 1.36 $\pm$ 0.09 * \\
$\alpha_{1}$ & 1.08 $\pm$ 0.06  &  0.95 $\pm$ 0.05 * \\
$\alpha_{2}$ &  1.39 $\pm$ 0.04 &  1.37 $\pm$ 0.26 \\
\hline
\end{tabular}
 * p value $<$ 0.05
\caption{The clinical and heart rate variables of subjects in the
study}
\end{table}

 The main findings of this study are compared with healthy controls.
Because of a relatively small patient population, the
results of this study do not allow us to draw conclusions regarding the lack of
the prognostic value of HRV in traditional measurements, 
but do give preliminary information
on the usefulness of fractal analysis methods in risk stratification of
patients with stable CHD.

\section{Conclusions}

Patients with stable angina pectoris had loss normal fractal
characteristics in heart rate variability estimated by non-linear
dynamic measures of heart rate behaviour. The measurement of a short-term 
fractal scaling exponent gives complementary information on
abnormal HR behaviour in patients with SAP in relation to other standard
measurements.
The present study shows that normal fractal properties of RR interval
dynamics are altered in patients with SAP. Dynamic analysis of HRV gives
independent information that probably cannot be detected by the traditional
linear analysis technique.
Healthy subjects have a distinct circadian rhythm of HRV, but this
rhythm seems to be blunted in coronary heart disease (CHD) patients.
Fractal correlation properties and fractal dimension in this study may
reflect an altered neuroanatomic interaction that may predispose patients 
to the development of CHD.
Further studies in a larger population will be needed to further define
the clinical utility of new fractal measurements of HRV for risk
stratification in patients with CHD.

\begin{figure}
\epsfig{file=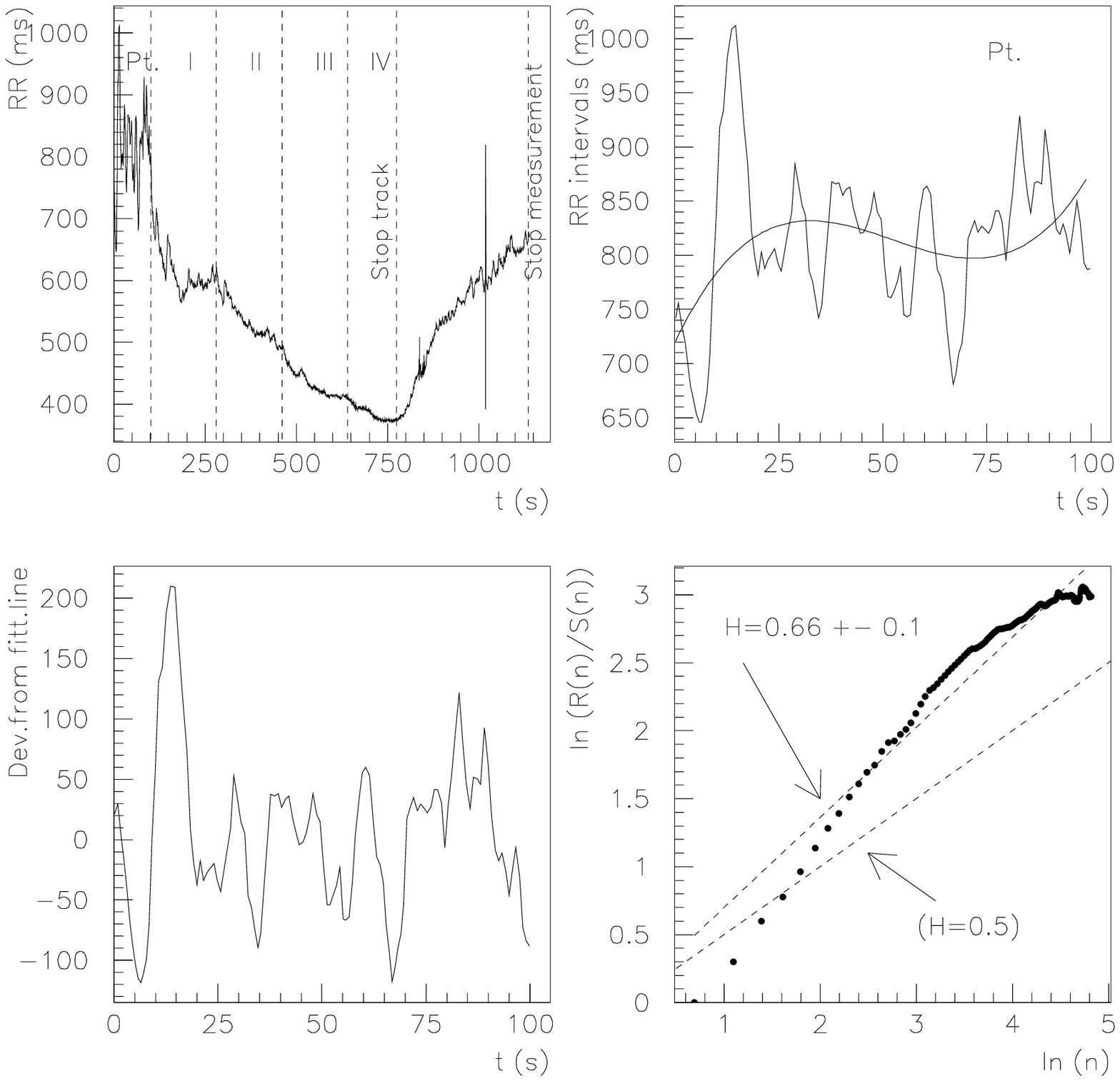,width=15.0cm}
\caption{Example of RR intervals in one typical ergometric measurement; RR
intervals in unforced regime (pretrigger, Pt.) with a cubic polynomial fit; 
deviations from the fitting curve; and the R/S calculation result of the 
Hurst exponent H in
comparison with random data, H=0.5.}
\end{figure}

\begin{figure}
\epsfig{file=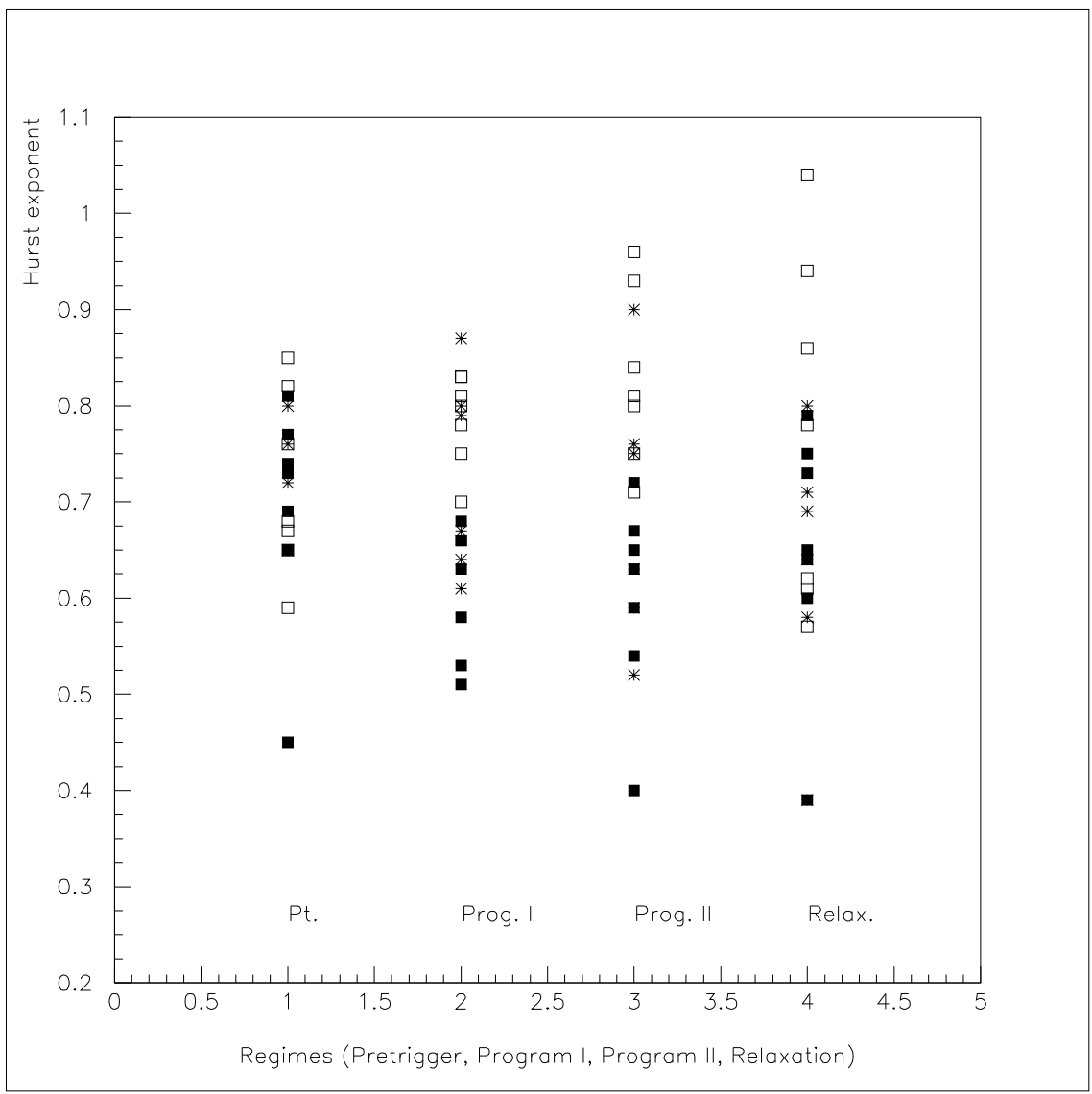,width=15.0cm}
\caption{Hurst exponents for different regimes in ergometric measurements;
Pretrigger (Pt.), Program I, Program II and Relaxation (30 sec., 3 min., 
3 min., 6 min. duration, respectively), with the cubic polynomial fit
before the R/S calculation. Open squares correspond to healthy subjects, 
filled squares to ill subjects
and stars to suspected subjects.}
\end{figure}

\begin{figure}
\epsfig{file=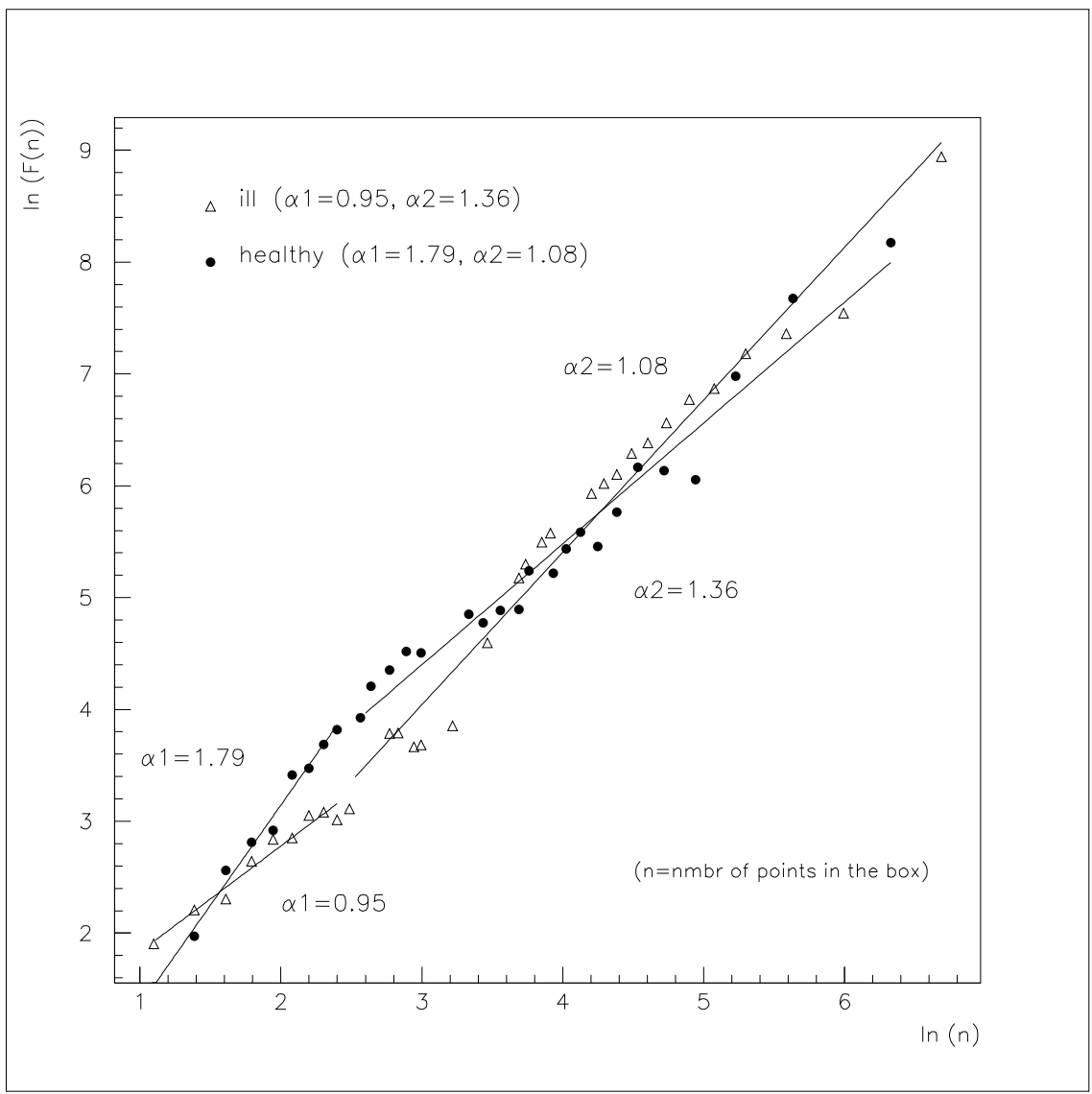,width=15.0cm}
\caption{RR intervals analysed using the DFA method showing the difference between healthy and
ill (SAP) subjects.}
\end{figure}

\end{document}